\documentclass[a4paper, conference]{IEEEtran}
\IEEEoverridecommandlockouts
\usepackage{cite}
\usepackage{amsmath,amssymb,amsfonts}
\usepackage{algorithmic}
\usepackage[linesnumbered,ruled]{algorithm2e}
\usepackage{graphicx}
\usepackage{textcomp}
\usepackage{xcolor}
\def\BibTeX{{\rm B\kern-.05em{\sc i\kern-.025em b}\kern-.08em
		T\kern-.1667em\lower.7ex\hbox{E}\kern-.125emX}}
	
\makeatletter
\def\ps@IEEEtitlepagestyle{%
	\def\@oddfoot{}%
	\def\@oddhead{\mycopyrightnotice}\relax
	\def\@evenhead{\@IEEEheaderstyle\thepage\hfil\leftmark\hbox{}}\relax
	\def\@evenfoot{}%
}
\def\mycopyrightnotice{%
	\begin{minipage}{\textwidth}
		\centering \scriptsize
		Copyright~\copyright~2023 IEEE.  Personal use of this material is permitted.  Permission from IEEE must be obtained for all other uses, in any current or future media, including reprinting/republishing this material for advertising or promotional purposes, creating new collective works, for resale or redistribution to servers or lists, or reuse of any copyrighted component of this work in other works. This paper is accepted at the 36th IEEE International Symposium on Defect and Fault Tolerance in VLSI and Nanotechnology Systems (DFT) 2023.
	\end{minipage}
}
\makeatother
	
\begin{document}

\title{On-Chip Sensors Data Collection and Analysis for SoC Health Management
\thanks{This work was supported in part by the European Union through European Social Fund in the frames of the "Information and Communication Technologies (ICT) programme" (“ITA-IoIT” topic) and by the Estonian Research Council grant PUT PRG1467 "CRASHLESS“.}
}

\author{\IEEEauthorblockN{Konstantin Shibin, Maksim Jenihhin}
\IEEEauthorblockA{\textit{Dept. of Computer Systems, Tallinn University of Technology}\\
Tallinn, Estonia \\
\{konstantin.shibin; maksim.jenihhin\}@taltech.ee}
\and
\IEEEauthorblockN{Artur Jutman, Sergei Devadze, Anton Tsertov}
\IEEEauthorblockA{\textit{Testonica Lab OÜ}\\
Tallinn, Estonia \\
\{artur; sergey; anton\}@testonica.com}
\and
\small This paper is a part of the Special Session \\"Towards cross-layer resilience: from reliability estimation at design phase to in-field error detection and on-chip sensor data processing."
}
\maketitle

\begin{abstract}
Data produced by on-chip sensors in modern SoCs contains a large amount of information such as occurring faults, aging status, accumulated radiation dose, performance characteristics, environmental and other operational parameters. Such information provides insight into the overall health of a system's hardware as well as the operability of individual modules. This gives a chance to mitigate faults and avoid using faulty units, thus enabling hardware health management.
Raw data from embedded sensors cannot be immediately used to perform health management tasks. In most cases, the information about occurred faults needs to be analyzed taking into account the history of the previously reported fault events and other collected statistics. For this purpose, we propose a special structure called Health Map (HM) that holds the information about functional resources, occurring faults and maps relationships between these. In addition, we propose algorithms for aggregation and classification of data received from on-chip sensors.
The proposed Health Map contains detailed information on a particular system level (e.g., module, SoC, board) that can be compiled into a summary of hardware health status that in its turn enables distributed hierarchical health management by using this information at a higher level of system hierarchy, thus increasing the system’s availability and effective lifetime.
\end{abstract}

\begin{IEEEkeywords}
SoC, MPSoC, on-chip sensor, health management, fault management, health map, self-awareness
\end{IEEEkeywords}

\section{Introduction}
Modern System-on-Chips (SoCs) and Multi-Processor System-on-Chips (MPSoCs) frequently contain a wide array of embedded on-chip sensors for monitoring the parameters and faults in the system. They are used during manufacturing for testing, measurement and adjustment of parameters, programming, etc. Increasingly, the embedded sensors and monitors (together called embedded instruments (EIs)) are used over the lifetime of the system\cite{TerBraak2010,Kochte2018}: for built-in self test (BIST), parameter monitoring (e.g., to estimate aging) and fault detection.
On the one hand, the data collected from EIs could be useful on its own (e.g., when a self-test has failed and the system must restart). On the other hand, when this data is collected and systematized, the statistics and trends might provide valuable insight into the nature of the problems as well as the remaining "health" of the system. 
This is especially important for complex systems with redundant resources where the health management (HM) approach can provide the possibility for graceful degradation, i.e., prolonging the lifetime of the system despite the failure of some of its parts. This approach becomes increasingly attractive for commercial off-the-shelf (COTS) components being used more in harsh environments (e.g., space radiation) \cite{Bokil2020,Esposito2015} due to higher performance, lower cost and better efficiency when compared to specialized components with classic protection mechanisms such as triple modular redundancy (TMR).
Moreover, when the collected data is stored, analyzed and summarized, it can be very useful in higher-level processes like the mapping of tasks to available resources to ensure graceful degradation. Another benefit is the possibility to provide the health status information to the higher levels in a hierarchical system.

EIs in the scope of this paper are embedded specialized modules built into the hardware of a SoC for collection of the service data related to the dependability of the system. For example, delay monitors can estimate the aging of a semiconductor IC. A parity checker or a property checker can detect incorrect operation (e.g., due to a bit flip resulting in a soft error). EIs can be very numerous and typically some test/service bus like IJTAG is used to access them \cite{autotestcon16,rsn-kochte-lats16}. This allows to decouple the EIs from functional resources like data buses, provide dedicated access to the EIs and implement additional features like triple-modular redundancy to increase the dependability or auxiliary asynchronous networks for fast fault detection \cite{natw14}.

While EIs can quickly and efficiently detect faults or other anomalies and measure parameters, the raw data from embedded sensors and monitors cannot be immediately used for the purposes like health management and prognostics. The collected data has to be analyzed and the faults have classified based on the context like previously collected statistics and the location of the fault. With EIs and corresponding access busses being separated from functional resources into a special layer, there is a need to describe and track the relationship between diagnostic and functional resources at the system level.
The accumulated health data must also be prepared to be useful for active fault mitigation. The health status of functional resources has to be available to scheduling algorithm for proper mapping of tasks to the available resources.

This paper details a special structure called Health Map (HM) that holds detailed information about functional resources, health information in terms of occurring faults and operating parameters and maps the relationships between these. In addition, we propose the algorithms for aggregation of the data received from EIs in order to compile a summary of hardware health status. This information can be then used at local level for task scheduling as well as at higher levels of system hierarchy for distributed hierarchical health management. This structure is a part of the overall health and fault management architecture (FMA) which is aimed at increasing the availability of an MPSoC-based system \cite{dnt17}.

\section{Related work}
Health data collection is implied by many works on the topic of self-aware system and health/resource management in SoCs, but it is barely discussed in detail.
Several works which are addressing on-chip fault handling ahd health/resource management do not consider historical health data, only relying on immediately detected faults \cite{TerBraak2010,Ibrahim2019}.
In many works some approach to collection of historical data about system health from embedded sensors and instruments is embedded into the concept of health management and self-awareness, however mostly it is only conceptually described \cite{Dutt2016,Guang2010,Kochte2018}.
Hierarchical agent framework as a monitoring layer for self-aware system is proposed in \cite{Guang2010}, but the authors do not detail the collection of health data required for consistent self-awareness.
Hierarchical health management is also applied to unmanned air vehicles (UAVs)\cite{Man2009,Ge2004}. In \cite{Man2009} the health management helps to evaluate the condition of UAVs and their sub-parts, diagnose faults and predict the remaining life, but the authors do not detail the accumulation of historical health data. 

\section{Health Map}\label{sec:health-map}

Raw data which is collected via embedded instrumentation cannot be immediately used to perform health management. In most cases, the information about occurred fault needs to analysed in the context of the previously reported fault events. For example, when a fault is detected by an instrument and reported through IJTAG network \cite{dnt13,natw14}, it is not possible to correctly classify it using only the data about that single fault event. In order to classify the fault, analysis of previous occurrences of the fault is essential.

For this purpose, a FMA maintains a centralized database, called \emph{Health Map} - HM, which holds the information about functional resources, occurring faults and the relationship between those two. HM preserves the statistics of fault occurrences, which can be used for fault classification and better prediction of resource reliability. This analysis and classification is performed by the software part of the FMA - the Fault Manager (FM) \cite{lats16}.

\begin{figure*}[tb]
	\centering
	\includegraphics[width=0.8\textwidth]{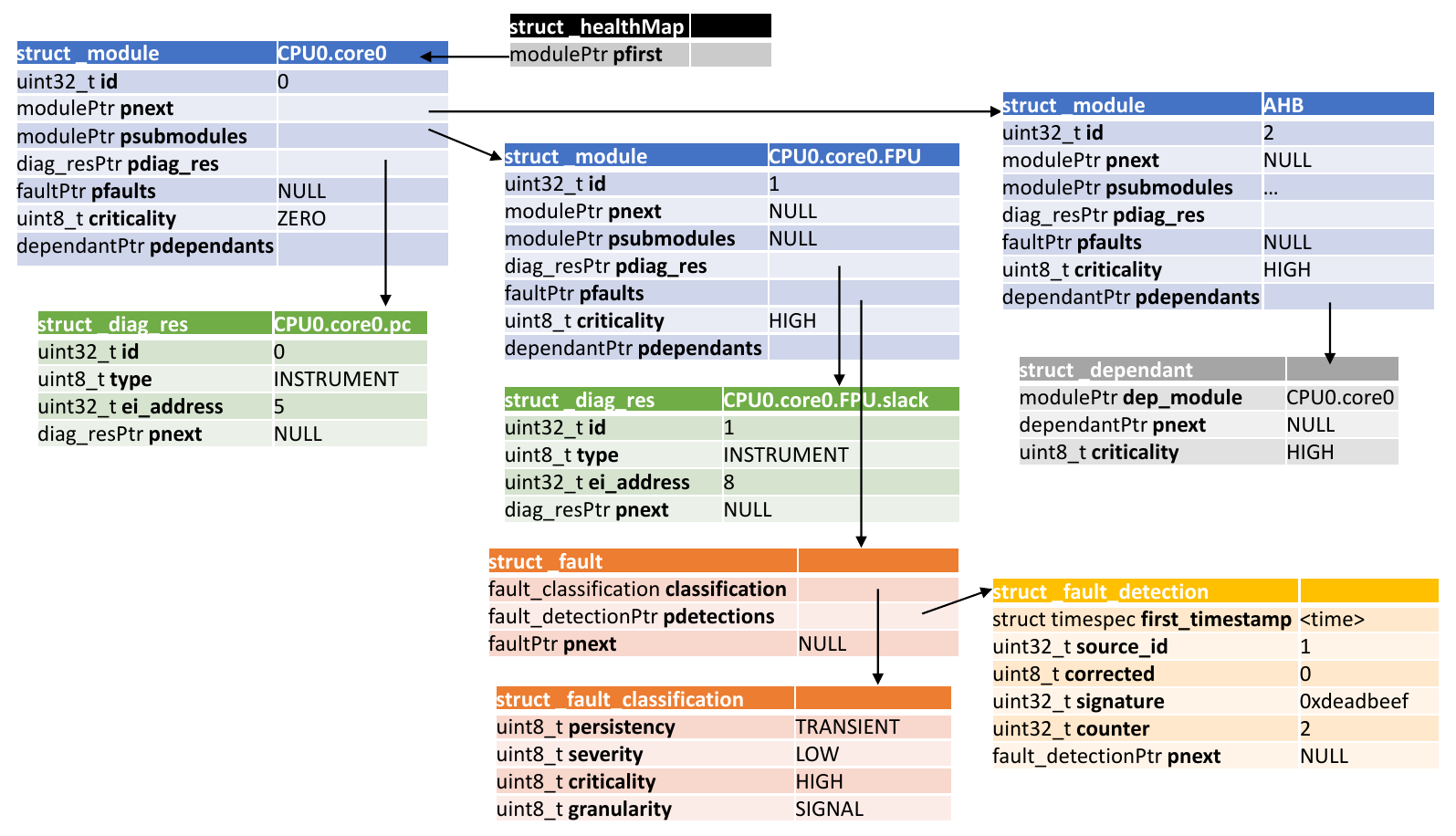}
	\caption[Health Map example data structure]{Health Map example data structure}
	\label{fig:health-map-data-structure} 
\end{figure*} 

Health Map is a collection of data structures that represent several different types of entities (described below). The connections between these entities define their actual relationships (e.g. fault detected in a functional resource). An array of similar entities is organized as a linked list. 

Figure~\ref{fig:health-map-data-structure} shows a simple example of Health Map consisting of the following entities: CPU cores with an FPU, faults and fault detection events.

The main types of entities (and their corresponding data structures) in HM are:

\begin{description}
	\item[Functional hardware resources (modules)] \hfill \\Fault manager must understand which system hardware resources the detected faults are related to. An initial map of all system resources should be provided together with the brand-new (undamaged) hardware.
	
	\item[Diagnostic resources] \hfill \\In the FMA, main bulk of information about faults is received from embedded instruments that monitor functional resources in hardware. HM describes this relationship. Hardware diagnostic resources always belong to a certain functional hardware resource in HM. Other diagnostic means (i.e. not related to embedded instrumentation) can also be supported potentially. For example, those are diagnostic methods based on SBST or BIST techniques.
	
	\item[Faults] \hfill \\Whenever a fault is detected, all relevant data must be captured and stored permanently in the HM. This ensures that faults, patterns of their occurrence (e.g. transient or intermittent) and other meta-data can be analyzed and extracted at any time later. Faults always belong to a certain functional hardware resource in HM.
	
	\item[Fault detections] \hfill \\Since one and the same fault can be detected several times and potentially by different diagnostic resources, the information about the detections is stored in an additional data structure. Fault detections always belong to a fault in HM.
	
	\item[Dependency] \hfill \\In some cases it is required to describe a relationship between two functional modules (system resources) in form of \emph{dependency}, i.e. when correct functioning of a module depends on operation of another module in a system. An example can be a peripheral connected to a certain bus: if a bus or its part fails, then this peripheral may become inaccessible (unusable). For such purposes, a separate data structure is introduced in HM. It provides a link to resource which is \emph{dependant} and the severity of that dependency.
\end{description}

\subsection{Preparation and run-time}
Health Map describes the functional and diagnostic resources in a specific system and the initial map must be prepared in advance based on the system's specification. This is done \emph{off-line} before deploying FMA to the target system. The proposed HM is initially described in XML format which is human-readable, but allows machine handling  and reuse of modules, for e.g. multiple instances of identical resources. This description is compiled into a binary form for efficient handling by FM and storage in memory. Serialized HM binary data structure is loaded to the non-volatile memory of the target system.

During \emph{run-time} FM code will handle the HM binary data structure with the initial mapping of functional and diagnostic and any existing faults. FM code and HM data must reside in protected memory (e.g. by error-correcting code (ECC)) at run-time in order to avoid health data corruption. FM code must be executed inside a "safety island" with dedicated error-protected processing cores e.g., in dual lock-step or TMR configuration. Many existing MPSoCs already have such "safety islands" with protected cores.
At start-up FM will validate the HM binary data structure and load it into RAM for fast access and modification. Upon addition of new fault data, this can be efficiently appended and written back to the non-volatile memory (see next subsection). The organization of HM in RAM allows efficient access for fault classification and preparing the summary of system's health (see Section~\ref{sec:resource-map}).

\subsection{Serialized Health Map}\label{sec:hm-serialized}
In order to preserve HM with accumulated fault data, it is necessary to store it in a non-volatile memory during system shutdown. Direct copying of data from RAM is not useful due to data structures with memory pointers, so a Serialized Health Map (SHM) version is used instead. SHM occupies subsequent uninterrupted memory region. This allows to save SHM as a binary file or store it in EEPROM memory. Similarly it can be read from a file or EEPROM and deserialized into run-time RAM data structures.

\begin{figure*}[tb] 
	\centering
	\includegraphics[width=0.8\textwidth]{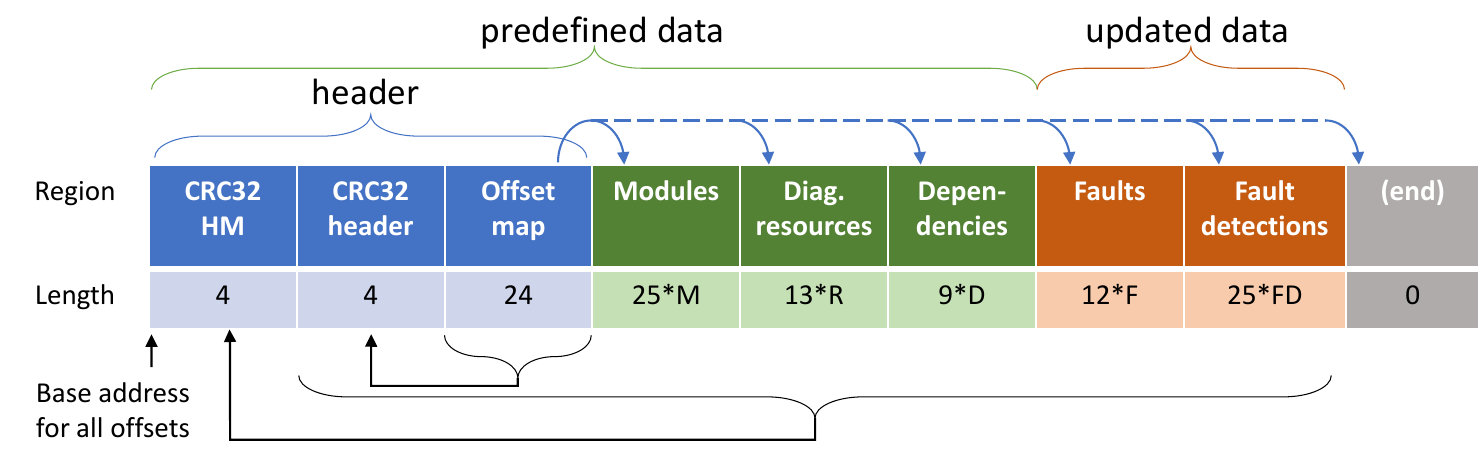}
	\caption[Serialized Health Map memory layout]{Serialized Health Map memory layout}
	\label{fig:health-map-layout} 
\end{figure*} 

SHM has specific layout in memory (shown in Figure~\ref{fig:health-map-layout}) that facilitates (de-)serialization and updating with new fault data. It is largely divided into two parts. In the \emph{predefined (constant)} part, the header and information about system's resources are prepared off-line. This part would be the same for all systems with identical hardware. \emph{Updated (dynamic)} part contains data about faults occurred throughout the system lifetime. This data is unique for each individual physical system. SHM is designed in a way that new data structures are appended to the end of SHM. This allows to minimize non-volatile memory erase and write operations. The header of SHM contains checksums (CRC32 algorithm) to ensure the integrity of the header as well as the whole data structure.

All memory offsets (shown as blue arrows above the layout structure in Figure~\ref{fig:health-map-layout}) are stored as absolute byte offsets using the beginning of HM as the base address. This ensures relocatability of HM: at any address in RAM, in a regular file, in an EEPROM, etc. During deserialization, these offsets are converted to usual memory pointers for fast access. The opposite is done during serialization procedure.

Length of each part of SHM is shown in Figure~\ref{fig:health-map-layout}. For the header it is fixed, while for other parts it depends on the number of entries of: \emph{M} - modules; \emph{R} - diagnostic resources; \emph{D} - dependencies; \emph{F} - faults; \emph{FD} - fault detections.

\subsection{Health Map pruning}
During prolonged system lifetimes and/or events which cause many faults to occur and to be detected, HM can accumulate a large number of fault detection entries. This will increase the amount of memory required for the HM data structure and the time required for statistical analysis of the fault data. In order to avoid accumulation of unnecessary data, two mechanisms are suggested for HM. Firstly, the \emph{fault detection} entry has \emph{counter} field which allows to register the number of similar events without creating a separate entry. Secondly, during system idle time, FM software can perform HM pruning and combine several fault or fault detection entries. This can be useful when, for example, an intermittent fault happens at distinct occasions and separate fault detection entries are created. It can be a policy of FMA to combine old fault or fault detection entries in order to reduce the HM and SHM memory footprints.

\section{Resource Map}\label{sec:resource-map}
Resource Map (RM) is a data structure in prepared by FM at run-time that holds the information about current health status of hardware resources. All modules defined in HM could be added to RM together with the information about worst severity and persistence of faults detected and attributed to corresponding modules (see example in Table~\ref{tab:rm-example}). This represents a quick summary of module's current health status and is used in calculating the mapping of software tasks to hardware resources.

\begin{table}[htb]
	\centering
	\caption[Resource map example]{Resource map example}
	\label{tab:rm-example}
	\begin{tabular}{l | p{1cm} | p{1.6cm} | c}
		Module name   & Worst severity & Worst persistence & Status \\
		\hline\hline
		CPU        & LOW  & TRANSIENT & PROPAGATED FAULT\\
		CPU.C0     & LOW  & TRANSIENT & PROPAGATED FAULT\\
		CPU.C0.FPU & HIGH & TRANSIENT & OWN FAULT       \\
		CPU.C1     & ZERO & ZERO      & AVAILABLE       \\ 
		CPU.C1.FPU & ZERO & ZERO      & AVAILABLE       \\ 
		CPU.C2     & ZERO & ZERO      & AVAILABLE       \\ 
		CPU.C2.FPU & ZERO & ZERO      & AVAILABLE       \\ 
		CPU.C3     & ZERO & ZERO      & MAINTENANCE     \\ 
		CPU.C3.FPU & ZERO & ZERO      & MAINTENANCE     \\ 
	\end{tabular}
\end{table}

\begin{description}
	\item[\emph{Module name}] \hfill \\refers to module name in HM.
	\item[\emph{Worst persistence}] \hfill \\stores worst persistence among all faults registered for the module.
	\item[\emph{Severity}] \hfill \\stores worst severity among all faults registered for the module.
	\item[\emph{Status}] \hfill \\indicates the current status of the module:
	\begin{description}
		\item[Available] \hfill \\Module is available, no faults registered.
		\item[Own fault] \hfill \\Module might have limited functionality due to a detected fault.
		\item[Propagated fault] \hfill \\Module might have limited functionality due to a fault detected in child modules.
		\item[Maintenance] \hfill \\Module is not available due to ongoing maintenance (BIST, SBST etc).
	\end{description}
\end{description}

\subsection{Resource map data}
In contrast to HM, Resource Map is available only during run-time. It is populated with data by scanning HM and searching for worst available persistence and severity levels among all faults registered for a given module. During calculation, criticality is also taken into account to calculate the propagation of faults to higher levels of module hierarchy (from child to parent).

After the initialization procedure RM may be updated whenever a new fault is detected. If its persistence and/or severity levels are worse than those in RM, the latter should be updated.

Each entry in RM corresponds to a module in HM and takes just 7 bytes of RAM: module ID takes 4 and enumerations (worst severity, worst persistence, status) take 1 byte each.

\subsection{Resource map update algorithms}\label{sec:rm-update-algorithm}

Algorithm~\ref{alg:rm-init} describes the operations needed for filling the RM. It populates RM at initialization by scanning through the whole HM for existing faults. In general, logic of the algorithm can be described as follows:

\begin{enumerate}
	\item For each module, iterate through all faults
	\item For each fault of a module, check if its severity and persistence is worse than previously found for the same module
	\item Write worst severity and persistence of the module to Resource Map
	\item If criticality of the module is not zero, calculate propagated fault to parent module and update its entry in Resource Map
\end{enumerate}

In the algorithms below, $s$ denotes severity: $s_{m}^{RM}$ is module's worst severity in RM and $s_f$ is the severity of a fault. The same index rules apply for $p$ (persistence) and $c$ (criticality). $m$ is a module having a set $F_m$ of faults $f$.

\begin{algorithm}[htb]
\SetKwFunction{propagateFault}{propagateFault}
\SetAlgoLined
\ForEach{$m \in RM$}{
	$s_{worst} = s_{m}^{RM}$\;
	$p_{worst} = p_{m}^{RM}$\;
	\ForEach{$f \in F_m$}{
		$s_{worst} = max(s_{worst}, s_f)$\;
		$p_{worst} = max(p_{worst}, p_f)$\;
	}
	$s_{m}^{RM} = s_{worst}$\;
	$p_{m}^{RM} = p_{worst}$\;
	$st_{m}^{RM} = own\_fault$\;
	\propagateFault{$m$}\;
	
}
\caption{Resource map initialization algorithm}\label{alg:rm-init}
\end{algorithm}

When a new fault is detected, RM needs to be updated using only the information of one given fault. This is done by \texttt{updateSingleFault} procedure listed below.

\begin{procedure}[htb]
	\KwIn{$m$ - module which receives a new fault}
	\KwIn{$s_f, p_f$ - severity and persistence of the new fault}
	\KwIn{$st_f$ - status of new fault in RM}
	$s_{m}^{RM} = max(s_{m}^{RM}, s_{f})$\;
	$p_{m}^{RM} = max(p_{m}^{RM}, p_{f})$\;
	$st_{m}^{RM} = st_f$\;
	\propagateFault{$m$}\;
\caption{updateSingleFault($m$,$s_f$,$p_f$,$st_f$)}\label{alg:rm-update-single}
\end{procedure}
Fault propagation procedure is called in both cases and its purpose is to take care of fault propagation through criticality of modules. The procedure is recursive as the fault can propagate to the parent of a parent module.
\begin{procedure}[htb]
	\KwIn{$m$ - module with faults to be propagated}
	\If{$c_m \neq zero$}{
		$s_{prop} = min(s_{m}^{RM}, c_m)$\;
		$m_p = parent(m)$\;
		\updateSingleFault{$m_p$,$s_{prop}$,$p_{m}^{RM}$,$prop\_fault$}\;
	}
\caption{propagateFault($m$)}\label{alg:fault-propagation-proc}
\end{procedure}

\subsection{Health-aware task scheduling}\label{sec:core-map}
CPU cores or \emph{processing cores} are part of system's hardware resources, but of special type because from the operating system (OS) perspective, they run the software tasks. In multi-core systems, several programs can be run in parallel and each of the latter is assigned to a certain processing core ID.

In the scope of FMA the faults are detected and attributed to certain modules in HM. If the accumulated fault(s) rendered the affected processing core unusable for some or all tasks, it is important that the OS scheduler does not schedule tasks to these cores anymore. It is essential that FM could translate a particular processing core module in HM to core ID understood by OS and instruct the OS scheduler not use certain cores for certain tasks.

This is achieved by comparing the status of processing cores in RM to the requirements of the tasks \cite{dnt17}. The requirements can include sub-modules like e.g., FPU. Based on the result of the comparison, FM fills a \emph{core affinity mask} for each task. Some OS schedulers (e.g. Linux) accept this information without modification of the scheduler code \cite{dnt17}.

\section{Hierarchical health management}\label{sec:hierarchical-hm}

\begin{figure}[tb]
	\centering
	\includegraphics[width=0.7\columnwidth]{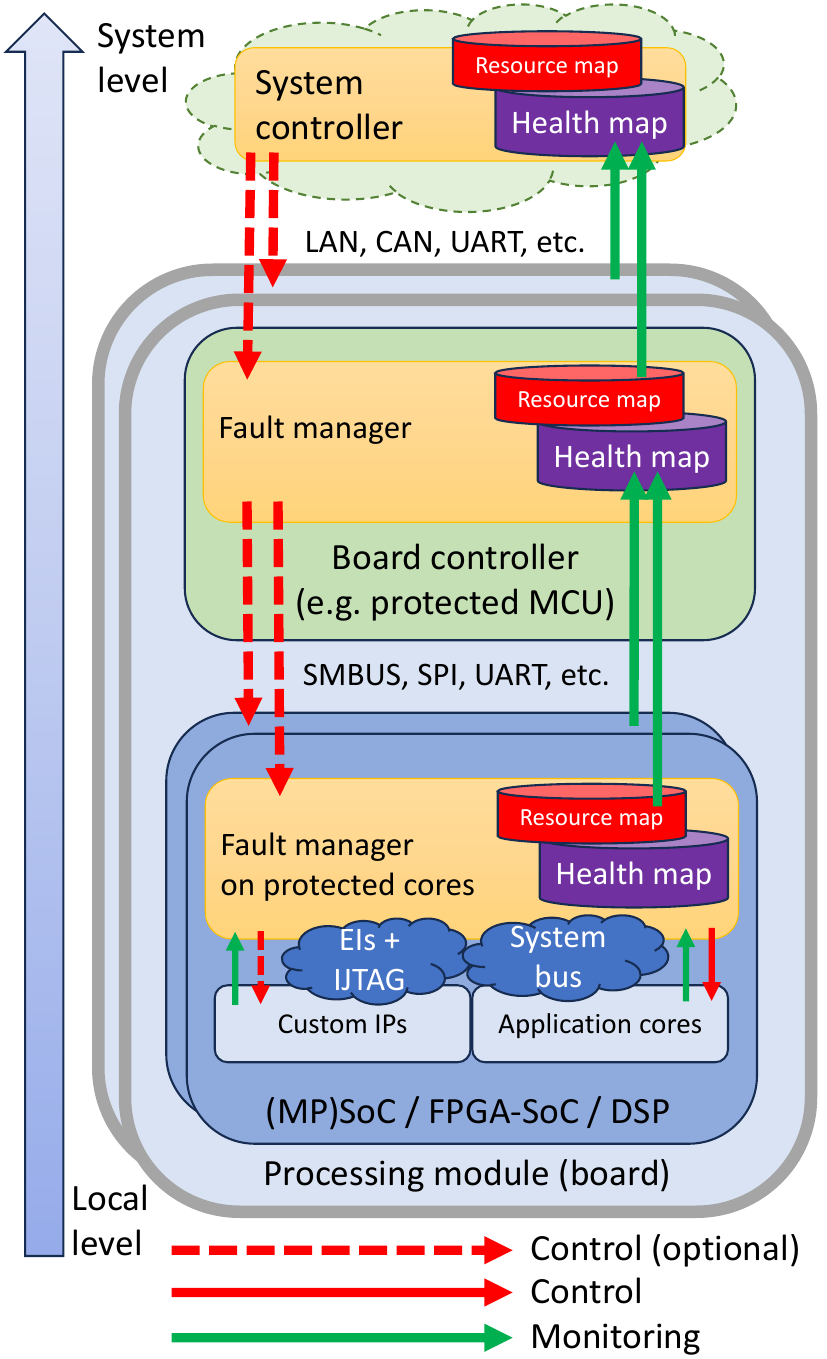}
	\caption[Health map data in an hierarchical system]{Health map data in an hierarchical system}
	\label{fig:hm-hierarchy} 
\end{figure}

Many systems consist of several layers of hierarchy. For example, an UAV would contain a data processing unit with an MPSoC, and on-board computer with a CPU and a board controller with a hardened microcontroller which monitors all other modules and sends telemetry data to ground. In such hierarchical system the health data should be summarized and collected from the lower levels (e.g. sub-core modules) to the upper levels (e.g. supervisor) in an hierarchical fashion, as shown in Figure~\ref{fig:hm-hierarchy}. Resource Map functionality fits very well into this picture by providing the summary of system's modules health ready to be sent to the higher level of hierarchy. The summarized health data from lower levels of hierarchy become the input for a Health Map at higher level of hierarchy. Data can be sent over any typical communication protocols in a serialized fashion similar to SHR.

At the same time, control commands (not discussed in this paper) are sent in the opposite direction, from higher to lower levels of system hierarchy, as the controller at higher (global) level has a better picture of overall system's health and can make high-level scheduling decisions accordingly. 

\section{Memory footprint estimation}\label{sec:hm-size}
Memory utilization of HM and RM data structures for a particular system will depend on the required number of entries of each type for storing the predefined and fault data. In this section we make an attempt to roughly estimate the number of entries in HM and RM for a typical system, as well as the amount of memory they occupy. The latter will be very similar for HM in RAM and SHM, except different headers (32 bytes in SHM vs 4 bytes in HM).

In an MPSoC, we expect that the size of HM would tend to correlate with the number of CPU cores and to have the assumed relationship of quantities of HM entities listed in Table~\ref{tab:estimation-modules}.

\begin{table}[htb]
	\centering
	\caption[Estimated size of HM for an MPSoC]{Estimated size of HM for an MPSoC}
	\label{tab:estimation-modules}
	\begin{tabular}{l | c | c}
		Entity & Entity type & Quantity\\
		\hline\hline
		CPU core & module & \emph{C}\\ 
		Core sub-modules & module & \emph{15C}\\ 
        Other modules & module & \emph{5C}\\
		System & module & \emph{10}\\
        \emph{Total (of the above)} & module & \emph{M = 16C + 10}\\
        \hline
		EIs & diag. resource & \emph{R = 3M}\\ 
        Dependencies & dependency & \emph{D = M}\\
        Faults & fault & \emph{F = 2D}\\
        Fault detections & f. detection & \emph{FD = 10F}\\
	\end{tabular}
\end{table}

For example, if we assume an MPSoC with 8 cores then the corresponding HM will have the quantities of data structures listed in Table~\ref{tab:estimation-mpsoc}.

\begin{table}[htb]
	\centering
	\caption[Estimated number of entities in HM for an MPSoC]{Estimated number of entities in HM for an MPSoC}
	\label{tab:estimation-mpsoc}
	\begin{tabular}{l | c | c | c}
		Entity type & Amount & Size (bytes) & Total (bytes)\\
		\hline\hline
        Header (SHM) & 1 & 32 & 32 \\
        Module & 138 & 25 & 3450 \\
        Diag. resource & 414 & 13 & 5382 \\
        Dependency & 138 & 9 & 1242 \\
        Fault & 276 & 12 & 3312 \\
        Fault detections & 2760 & 25 & 69000 \\
        \hline
        \textbf{Total} & & & \textbf{82418}
	\end{tabular}
\end{table}
The amount of memory required for RM (in case all modules are included) is: \emph{138 modules * 7 bytes = 966 bytes}.

This estimation shows that even for a fairly large MPSoC the amount of memory required by the corresponding HM with fault data is in the order of 1E5 bytes, for RM it is in the order of 1E3 bytes. This shall easily fit into RAM and persistent memory of modern MPSoC-based systems. 

\section*{Conclusion}
In this paper we present and detail Health Map, a data structure and related algorithms which play important roles in health and fault management architecture for SoCs. Fault data from on-chip sensors must be collected and analysed in order to accumulate historical data for accurate fault classification and health-aware resource usage. We show that the proposed Health Map is well suited for health management in modern SoCs by collecting and summarizing important information about SoC health. This information can be used for health-aware task mapping by OS as well as aggregation in higher levels of system hierarchy. An estimation of memory footprint shows that even for complex SoCs the overhead is minimal.

\bibliographystyle{IEEEtran}
\bibliography{references}

\end{document}